\documentclass[journal]{IEEEtran}
\usepackage{ifpdf}

\ifpdf
\pdfoutput=1\relax                   
\usepackage{graphicx}                
\DeclareGraphicsExtensions{.pdf,.png,.jpg,.jpeg} 
\else
\DeclareGraphicsExtensions{.eps}     
\fi%

\graphicspath{{figures/}{pictures/}{images/}{./}} 

\usepackage{microtype}                 
\PassOptionsToPackage{warn}{textcomp}  
\usepackage{textcomp}                  
\usepackage{mathptmx}                  
\usepackage{times}                     
\usepackage{cite}                      
\usepackage{tabu}                      
\usepackage{booktabs}                  

\usepackage{tabularx,ragged2e,booktabs,caption}
\usepackage{bm}
\usepackage{enumerate}
\usepackage{indentfirst}
\usepackage{multirow} 
\usepackage{rotating} 
\usepackage{booktabs} 
\usepackage{setspace} 
\usepackage{color}   
\usepackage{verbatim}
\usepackage{amsmath}

\usepackage{hyperref}
\usepackage{amssymb}
\usepackage{amsfonts}
\usepackage{floatrow}
\usepackage{bm} 
\usepackage[locale=FR]{siunitx} 
\usepackage[font=footnotesize,labelfont=sf,textfont=sf]{caption}
\usepackage{paralist} 
\usepackage[square,sort,comma,numbers]{natbib}

\usepackage{natbib}
\usepackage{algorithm}
\usepackage{algorithmic}
\usepackage{verbatim}

\usepackage[locale=FR]{siunitx} 

\usepackage{scalerel} 

\usepackage{multicol}
\ifCLASSINFOpdf
\else
\fi
\begin{document}
%
\title{Vector Field Streamline Clustering Framework \\
	for Brain Fiber Tract Segmentation}
%
%
%

\author{Chaoqing Xu, Guodao Sun, Ronghua Liang, Xiufang Xu
	
\thanks{C. Xu, G.Sun, and R. Liang are with Zhejiang University of Technology.
Hangzhou 310023, China. e-mail: \{superclearxu, godoor.sun\}@gmail.com., rhliang@zjut.edu.cn.}
\thanks{X. Xu is with Hangzhou Medical College. E-mail: 2659189077@qq.com} }
\maketitle

\begin{abstract}
Brain fiber tracts are widely used in studying brain diseases, which may lead to a better understanding of how disease affects the brain. The segmentation of brain fiber tracts assumed enormous importance in disease analysis. In this paper, we propose a novel vector field streamline clustering framework for brain fiber tract segmentations. Brain fiber tracts are firstly expressed in a vector field and compressed using the streamline simplification algorithm. After streamline normalization and regular-polyhedron projection, high-dimensional features of each fiber tract are computed and fed to the IDEC clustering algorithm. We also provide qualitative and quantitative evaluations of the IDEC clustering method and QB clustering method. Our clustering results of the brain fiber tracts help researchers gain perception of the brain structure. This work has the potential to automatically create a robust fiber bundle template that can effectively segment brain fiber tracts while enabling consistent anatomical tract identification.
\end{abstract}


%
\IEEEpeerreviewmaketitle



\section{Introduction}

\IEEEPARstart{D}{iffusion} tensor image (DTI) is an advanced magnetic resonance image (MRI) technique that enables radiologists an intuitive understanding of biological tissues in one’s brain\cite{pajevic2003parametric}. To help neuroscientists with an in-depth analysis of the brain functional regions and anatomical structures, scholars have developed brain fiber reconstruction and tracking methods based on DTI to generate fiber tracts, which illustrate biological connectivity in one’s brain\cite{TOURNIER2019116137,SMITH20121924}. Each brain fiber is presented as a streamline simulating the water distribution trend of a number of neurons. With thousands of brain fiber tracts constructed in a brain, biologists find it a great impediment to the analysis of inner brain connection. In quantitative analysis of brain connectivity, cortical-parcellation-based methods and clustering-based methods are used in brain segmentation\cite{o2013fiber}. Cortical-parcellation-based methods generally leverage a brain atlas, aiming at brain connectivity among different brain cortical regions. However, these methods sensitive to the ends of fiber tracts and are not conducive to understanding of fiber bodies. Complementary, clustering-based method group white matter fiber tracts based on fiber tract bodies, focuses on white matter anatomy segmentation with high consistency.

Brain fiber tracts are typicaly regarded as spatial streamlines. Fiber clustering methods can be separated into two categories according to similarity calculations: geometry-based approaches and parcellation-based approaches. Geometry-based approaches generally regard brain fiber tracts as geometric curves and compute the spatial distance among pairwise fiber tracts\cite{kumar2017fiberprint, o2013fiber, siless2018anatomicuts}. Based on the distance matrix, researchers can effectively explore the similarity of brain fiber tracts. While applying conventional clustering algorithms, such as hierarchical clustering, DBSCAN, and K-Means, researchers can obtain brain fiber tracts of each category. In general, such a method would inevitably lead to a high computational issue, especially working on large-scale datasets. Scientists also introduce mechanisms to speed up distance calculation\cite{garyfallidis2012quickbundles, VAZQUEZ2020117070}. However, such mechanisms may lead to misunderstanding of the physiological structure of brain fibers without anatomical guidance. Parcellation-based approaches have been proposed to increase physiological information, which would work for solving the aforementioned issue. Such methods generally use brain atlas centers or cortical parcellations to guide streamline similarity calculation\cite{wassermann2016white, bassett2017small,ingalhalikar2014sex}. Although parcellation-based clustering is popular in brain structure identification and anatomical connectivity analysis, it does not always divide the brain into recognizable bundles of fibers. Additionally, with the the improvements in deep learning in medical image analysis, fiber clustering based on deep learning approaches begins to show off. Current research on this topic is relatively rare. Existing works transform brain fiber tracts into topological function by using 3D brain surface\cite{gupta2017fibernet,gupta2018fibernet}. However, the volumetric parameterization of brain fibers highly relies on the shape-center of the MRI template image but lacks the capability of clustering brain fibers at multiple levels.

In this paper, we propose a vector field fiber tract clustering framework that can effectively cluster brain fibers at multiple levels. This approach is data-driven and highly relies on global fiber shape and fiber spatial positions. The central idea is to present brain fibers as vector field streamlines and construct features for each of the streamlines to avoid pairwise streamline distance calculation. First, we introduce a vector field streamline simplification method. Each brain fiber is normalized to the same length by applying the streamline simplification algorithm. Second, the simplified streamlines are projected onto the planes of a regular polyhedron. Third, we compute 3D space features and project plane features for each streamline. Thus, the vector field brain fibers are transformed into a high-dimensional feature space. Fourth, with the advanced deep embedded clustering algorithm (IDEC)\cite{ijcai2017-243}, brain fibers can be divided into multiple categories, which can be explained to a certain extent in physiology. Our contributions are as follows:
\begin{itemize}
	
\item A fiber tract clustering framework that consists of fiber simplification, regular-polyhedron projection, fiber feature construction, and deep clustering.

\item A vector field fiber simplification algorithm used to compress brain fiber data while maintaining fiber shapes. 

\item Various types of fiber feature construction methods that can be used to extract the spatial properties of streamlines adequately.

\item An adaptive improvement of the IDEC clustering method and experimental analysis on brain fiber data.

\item Qualitative and quantitative comparison between our clustering methods and QB clustering method using real data. 
	
\end{itemize}


\section{Related Work}

\subsection{Streamline Simplification Methods}
Streamline simplification is a general problem in many research fields,such as, image processing and geometry. A number of works have been reported on this topic. Carmo et al. \cite{10.1007/978-3-540-30135-6_55} presented a velocity flow simplification method based on hierarchical clustering with local linear expansion employed to extract critical topological flow features. Critical points play an important role in local linear expansion, which might lack a comprehensive understanding of complicated flow trends. Buchin et al. \cite{DBLP:journals/corr/abs-1806-02647} presented a progressive geometric curve simplification method for a different level of detail by using a shortcut graph based on Hausdorff distance to speed up the algorithm. This method works efficiently with progressive and non-progressive problems. Shen et al. \cite{shen2018new} provided a line simplification method based on image processing using raster data and performs well on unstructured image data. The Douglas–Peucker (DP) algorithm was developed using tolerance distance to divide the streamline with anchor and floater recursively changed to eliminate unnecessary points. Wang and Muller \cite{doi:10.1559/152304098782441750} proposed a line simplification method using cartographic and geographic characteristics and reported its good performance on large-scale areas with line boundaries. Jiang and Nakos \cite{jiang2003line} presented an approach to line simplification based on geometry detection that bottom on vector quantization and vector projection, which relies on self-organizing maps. Dyken et al. \cite{Dyken2009} developed a triangulation-based curve simplification method that can preserve the spatial relationship between adjacent curves. Park et al. \cite{PARK20111267} performed a hybrid line simplification method that combines quantitative characterization by using multiple-line simplification algorithms and line segmentation versus simplification based on previous characteristics. A near-linear approximation algorithm, which is based on Frechet error measures, was developed for simplifying high-dimensional curves \cite{10.1007/978-3-540-30135-6_55}. The experimental results demonstrate the efficiency of the algorithm, however, the evaluation of the simplification results has not been offered. 

\subsection{ Unsupervised Deep Clustering }
The concept of "Deep Clustering" firstly appeared in a conference paper\cite{hershey2016deep}, in which researchers introduced a deep learning framework for acoustic source segmentation and separation. This concept becomes increasingly influential in recent years and has developed a series of branches. Unsupervised deep clustering is one of the categories of deep clustering. Deep Belief Networks (DBN) is one of the pioneers in unsupervised deep learning, which is developed for hierarchical representations from unlabeled images\cite{10.1145/1553374.1553453}. Inspired by the unsupervised feature learning with DBNs, Chen\cite{chen2015deep} developed a deep belief network with nonparametric clustering (DNC), which performs a discriminative clustering model under the maximum margin framework and finetunes the parameters in the DBNs. Following DBN, a discriminatively boosted clustering method (DBC) has been proposed for image analysis\cite{li2018discriminatively}. DBC first extracts image features through fully convolutional auto-encoders and adopts a boosted discriminative distribution for training a kmeans clustering model. With almost the same architecture, deep embedded clustering DEC\cite{DEC} becomes one of the most representative unsupervised deep clustering. In contrast to DBC, DEC takes a deep autoencoder as the network structure and drops the decoder layers. The output of the encoder layers serves as the input of the standard k-means clustering module. The clusters are iteratively optimized by minimizing a KL-divergence of the clustering objectives. The performance of DEC shows significant robustness in unsupervised learning tasks. Improved from DEC, improved deep embedded clustering (IDEC) retains the decoder layers of autoencoder and preserves the reconstruction loss of the deep learning architecture\cite{ijcai2017-243}. By integrating the reconstruction loss and the clustering loss, IDEC can maintain the local structure of data and prevent the distortion of embedded space. The empirical results of IDEC in image and text datasets demonstrate the structure preservation. In addition, an advanced unsupervised online deep learning method (ODC) has been proposed to solve the unstable learning of visual presentation in the training schedule\cite{zhan2020online}. Performance and stability have been demonstrated through the evaluation of ImageNet data and Places data. 

\begin{figure*}[h]
	\centering
	\includegraphics[width=7in]{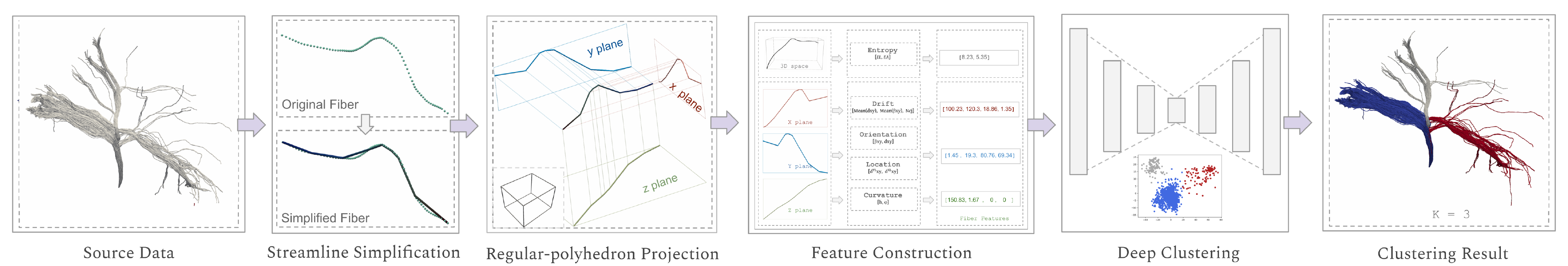}
	\caption{ Brain fiber clustering pipeline. Starting from the source data in the vector field, we first simplify the fiber tracts by using the streamline simplification algorithm, then we decompose it to different planes through regular-polyhedron projection. After the feature construction of each streamline, we obtain the feature matrix of the brain fibers and feed it to a deep clustering algorithm. The clustering results are then rendered helping researchers better visualize the brain fibers.}
	\label{fig:pipeline}
\end{figure*}

\begin{figure*}[h]
	\centering
	\includegraphics[width=7in]{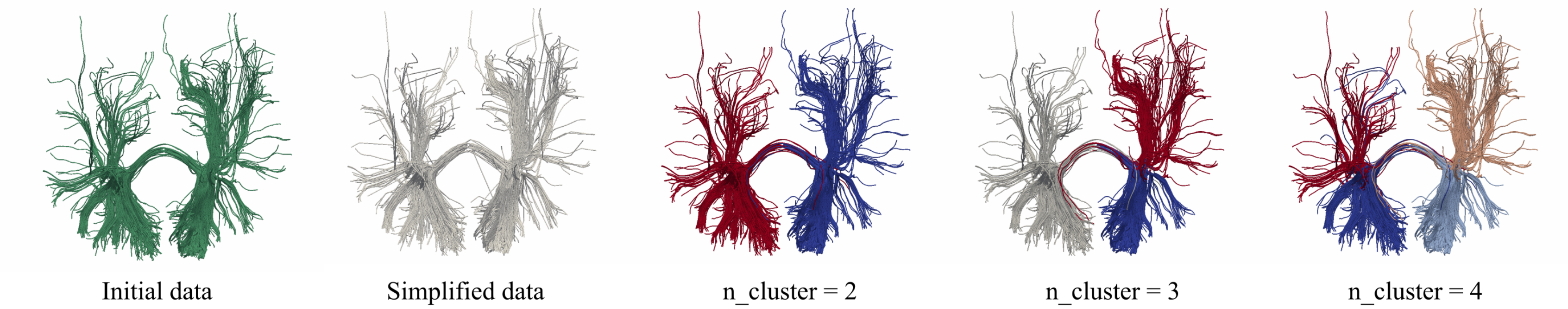}
	\caption{Our clustering of brain fiber data. It shows the rendering of initial brain fiber data, the simplified brain fiber data, and the clustering results of the brain fiber data.}
	\label{fig:n_clusters}
\end{figure*}

\subsection{ Brain Fiber Clustering }
According to the mechanisms of clustering methods, fiber clustering can mainly be divided into two categories: geometry-based\cite{kumar2017fiberprint, o2013fiber, siless2018anatomicuts,guevara2012automatic,garyfallidis2018recognition} and parcellation-based\cite{wassermann2016white, bassett2017small,ingalhalikar2014sex,yeh2016connectometry,8964333}. A typical framework for geometry-based fiber clustering defines similarity/distance between pairwise fiber tracts. This framework involves many similarity matrices, including Euclidean distance\cite{xu2020visual}, Hausdorff distance\cite{JIN201475}, Mahalanobis distance\cite{MADDAH2008191},etc. 
Román\cite{roman2017clustering} proposed a whole-brain fiber clustering method based on distance metric, which is computed by Euclidean distance between corresponding points of two fibers. Meanwhile, researchers also provided an efficient clustering method for large-scale brain fiber dataset\cite{VAZQUEZ2020117070}. The method maintains a good similarity between fiber shapes by using a restrictive flipped distance and reduces time complexity that makes it capable to handle millions of fibers. QuickBundles (QB) utilizes direct, flipped and minimum average direct-flip distances to compute the pairwise distance of fiber tracts rapidly\cite{garyfallidis2012quickbundles}. The results show that their approach can overcome the complexity of a large brain fiber dataset and efficiently provide robust clustering results. Apart from the computing efficiency of brain fiber clustering, parcellation-based methods aim at extracting brain fiber clusters that can be described anatomically. Such methods generally use brain atlas to guide fiber tract clustering. A recent study provided an automatically annotated fiber cluster method for whole-brain white matter structure identification and annotation through multiple subjects\cite{wu2018investigation}, and reproducibility was evaluated by multiple large-scale datasets\cite{zhang2019test}. Researchers provided an unsupervised parcellation-based clustering method that uses a publicly available atlas as anatomically-aware information\cite{wassermann2010unsupervised}. The clustering framework uses Gaussian process representation for each fiber, which spans a metric space and helps combine fiber bundles to better compute fiber similarity. Jin et al.\cite{JIN201475} presented an automatic clustering of brain fibers combing ROI-based clustering and distance-based clustering. This method first applies ROI constraints to brain fibers and then define fiber distance metric by using symmetric Hausdorff distance. In addition, with the prosperity of deep learning applications, researchers start working on deep learning which is the first approach for brain fiber tracts clustering using deep learning technique\cite{gupta2017fibernet}. The framework first transforms brain fiber tracts into a topological function and then uses convolutional neural networks to learn the shape features of brain fibers. The improved version of FiberNET\cite{gupta2018fibernet} reduces the effects of false-positive fibers. The reliability of the framework is evaluated through a series of testing on real data. 

\section{Workflow}

We begin with an overview of our framework before going through the detailed description of each step. Fig.~\ref{fig:pipeline} shows a schematic of our proposed workflow. The entire fiber clustering process consists of four steps: vector field streamline simplification, regular-polyhedron projection, streamline feature construction, and deep clustering. Given that brain fiber tracts generally appear as spatially dense streamlines, brain fiber tracts should be compressed and the spatial shapes should be maintained. Thus, we first present brain fiber tracts as vector field streamlines, and then perform vector field streamline simplification algorithm to obtain simplified fiber tracts. Each fiber tract would be normalized to the same length. Afterward, we project the fiber tracts to each plane of a regular polyhedron. By decomposing brain fiber tracts into multiple spatial planes, brain fiber spatial properties can be described more comprehensively. In this trial, we use a regular hexahedron for fiber tract decomposition and prove the effectiveness of experiments. By employing feature construction algorithms, we construct features in primitive space and on regular polyhedron planes for each of the brain fibers. Five types of features are constructed to reveal streamline properties. After then, we employ a deep clustering algorithm to classify the fiber tracts. Finally, with the contrastive color mapping (each fiber category is colored in a different color), the classified fiber tracts are rendered for intuitive visual comparison.

Before elaborating on each step, Fig.~\ref{fig:n_clusters} gives an example of our clustering results of brain fiber data. The initial brain fiber data (2467 streamlines) are rendered as path tubes to produces high-quality visualization with an enhanced spatial perception that better shows the brain fiber structure. The second panel shows the simplified brain fiber tracts using the simplification parameters ($\theta = 30$ and $L = 16$). Each fiber tract is discretized to have $16$ nodes and is used to construct brain fiber features. The third panel shows the two clusters of the brain fiber tracts, by using the deep clustering algorithm. We can clearly see the brain fibers are split into the left and right sides (the left brain and the right brain). Moreover, the third panel and the last panel show the brain fiber tracts that are classified into more categories, which are rendered with distinct colors. The clustering is consistent and each category of brain fiber tracts is spatially independent. This strategy is useful for interactive application because it maintains consistent clustering and allows intuitive visual comparison of small segmentation of brain fiber tracts. We describe each step in detail in the next four sections.

\begin{figure}[h]
	\centering
	\includegraphics[width=2.6in]{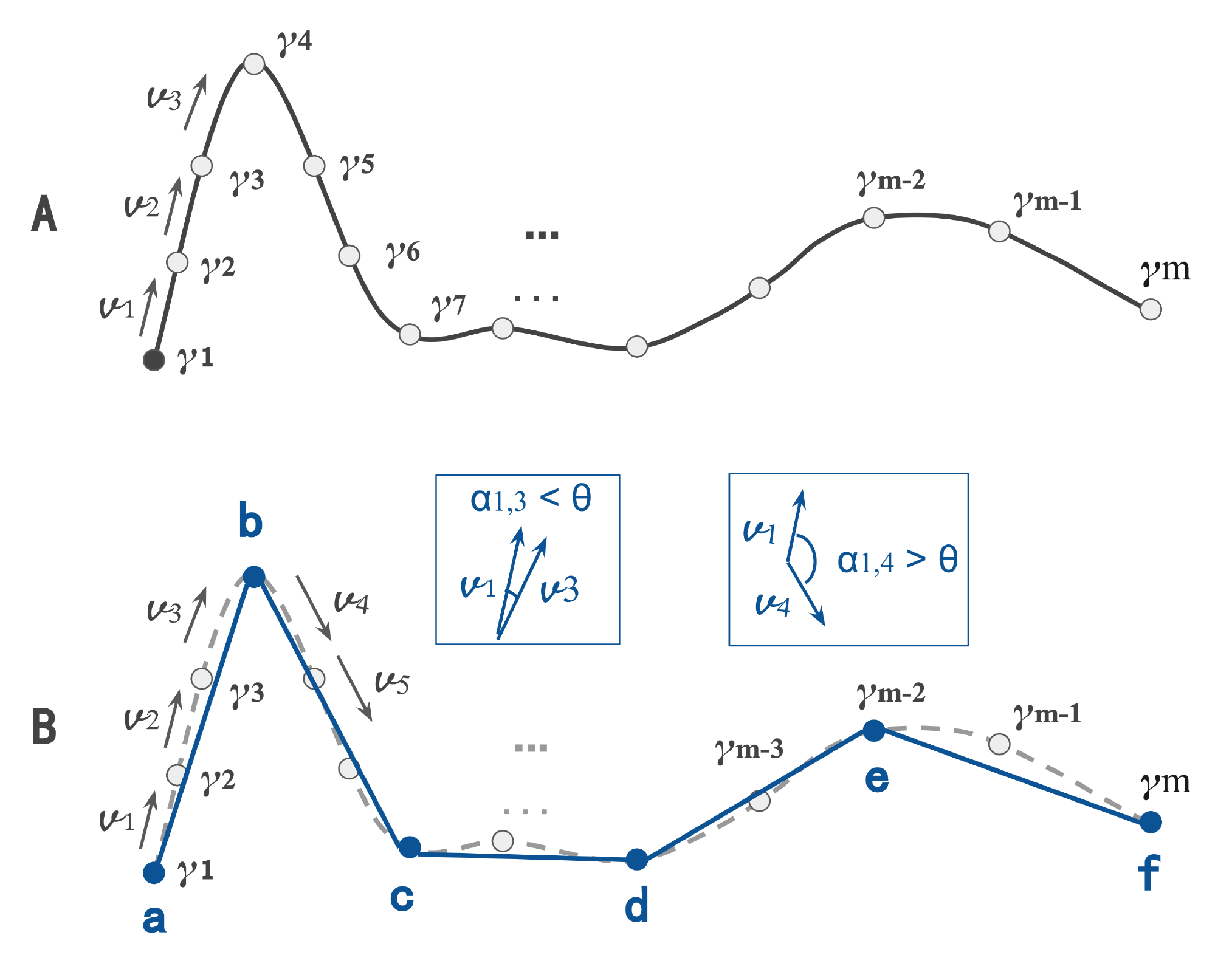}
	\caption{Illustration of streamline simplification. (A) shows a vector field streamline. The streamline consists of $m$ nodes starting from \begin{math}\gamma^1\end{math} to \begin{math}\gamma^m\end{math}. (B) indicates the streamline simplification method. $a$ and $f$ are the first node and the last point of the streamline.  $b$, $c$, $d$, and $e$ are the key nodes of the streamline, where the corresponding node indices are $s$, $u$, $v$, and $w$. The key node angles (\begin{math}\alpha_{1,s}\end{math}, \begin{math}\alpha_{s,u}\end{math}, \begin{math}\alpha_{u,v}\end{math} and \begin{math}\alpha_{v,w}\end{math}) are larger than \begin{math}\theta\end{math}. }
	\label{fig:fiberSimplification}
\end{figure}

\section{Streamline Simplification}

A core point of the streamline simplification method is to determine which nodes should be kept as important ones to represent streamline changes while omitting other nodes. A brain fiber tract is usually described as a vector field curve, which is composed of a series of vectors (Fig.~\ref{fig:fiberSimplification}(A)). The angle between two adjacent vectors can be used to describe how streamline fragments change locally. Two cases about the angle should be considered: first, a streamline changes sharply, where the angle between adjacent vectors is extremely large. In this case, a streamline appears in an abrupt shape. The turning nodes of the streamline can be determined as key nodes, and the corresponding vector is regarded as key vectors. Point $b$ in Fig.~\ref{fig:fiberSimplification}(B) shows such a key node. Second, a streamline changes gradually, in which the angle between adjacent vectors shows small values, and the streamline presents as a gentle curve. In this case, two pints with a long distance in between would be considered as key nodes. Point $c$ and node $d$ are two nodes in this case. According to this observation, we present a streamline simplification method. 

A brain fiber tract can be regarded as streamline that consists of a series of successive nodes: \begin{math}\gamma^1\end{math}, \begin{math}\gamma^2\end{math}, ... \begin{math}\gamma^i\end{math} ..., \begin{math}\gamma^m\end{math}, where the corresponding vectors are $ \vec{\bm{v_1}}, \vec{\bm{v_2}}, ... \vec{\bm{v_i}} ... \vec{\bm{v_{m-1}}} $ in the vector field (Fig.~\ref{fig:fiberSimplification}(A)).  $\vec{\bm{v_i}}$ can be figured out by adjacent two nodes when $i$ is not $1$ and $m$ and the node \begin{math}\gamma^i\end{math} = ($x_i$, $y_i$, $z_i$): 
\begin{equation}
	v_i = \frac{\gamma^{i+1} - \gamma^i}{\| \gamma^{i+1} - \gamma^i \|} = (\bar{x_i}, \bar{y_i}, \bar{z_i})
\end{equation}

In this paper, we define \begin{math}\alpha_{i,j}\end{math} to represent the angle between vector $\vec{\bm{v_i}}$ and vector $\vec{\bm{v_j}}$, where $ 0< $ $\bm{i}$, $\bm{j}$ $<m $: 
\begin{equation}
	 \alpha_{i,j} = \frac{180}{\pi} \times \frac{ \vec{\bm{v_i}} \cdot \vec{\bm{v_i}} }{ \left| \vec{\bm{v_i}} \right| \times \left| \vec{\bm{v_j}} \right| }
\end{equation}

Moreover, we introduce \begin{math}\theta\end{math} as a parameter determining streamline curving similarity (maintain streamline shape). The nodes with angle smaller than \begin{math}\theta\end{math} would be omitted. Larger \begin{math}\theta\end{math} indicates higher degree of simplification. The simplification steps of a streamline are as follows: considering that the first and last nodes of a fiber tract are generally located on the brain cortical surface revealing the connection of the brain cortex, we take both of them as key nodes. We then figure out the key nodes of the body of a streamline that is illustrated in Fig.~\ref{fig:fiberSimplification}B. Starting from the first node $a$ (the first key node), we compute the angle between key vector $\bm{v_1}$ and subsequent vectors until the angle value reaches \begin{math}\theta\end{math} (\begin{math}\alpha_{1,s-1}\end{math} $<$  \begin{math}\theta\end{math} and \begin{math}\alpha_{1,s}\end{math} $>$  \begin{math}\theta\end{math}). Thus, we regard point $b$, the corresponding vector is \begin{math}\alpha_{1,s}\end{math}, as an key point. Repeat this process until it reaches the last node of a streamline. Since each streamline should be normalized to the same length, the other parameter for streamline simplification is length ($L$), which means how many key nodes should be reserved. It contributes to streamline compression and normalization. The influence of the two parameters on streamline simplification is not independent. Small $L$ value leads to high streamline compression but low curving similarity, while a large $L$ value causes low streamline compression but high curving similarity. Striking a balance between these parameters would optimize the streamline simplification effect. In section \ref{ParameterDetermination}, we will discuss and evaluate the impact of the two parameters in-depth and determine their values in a specific application scenario.

\begin{figure}[h]
	\centering
	\includegraphics[width=2.6in]{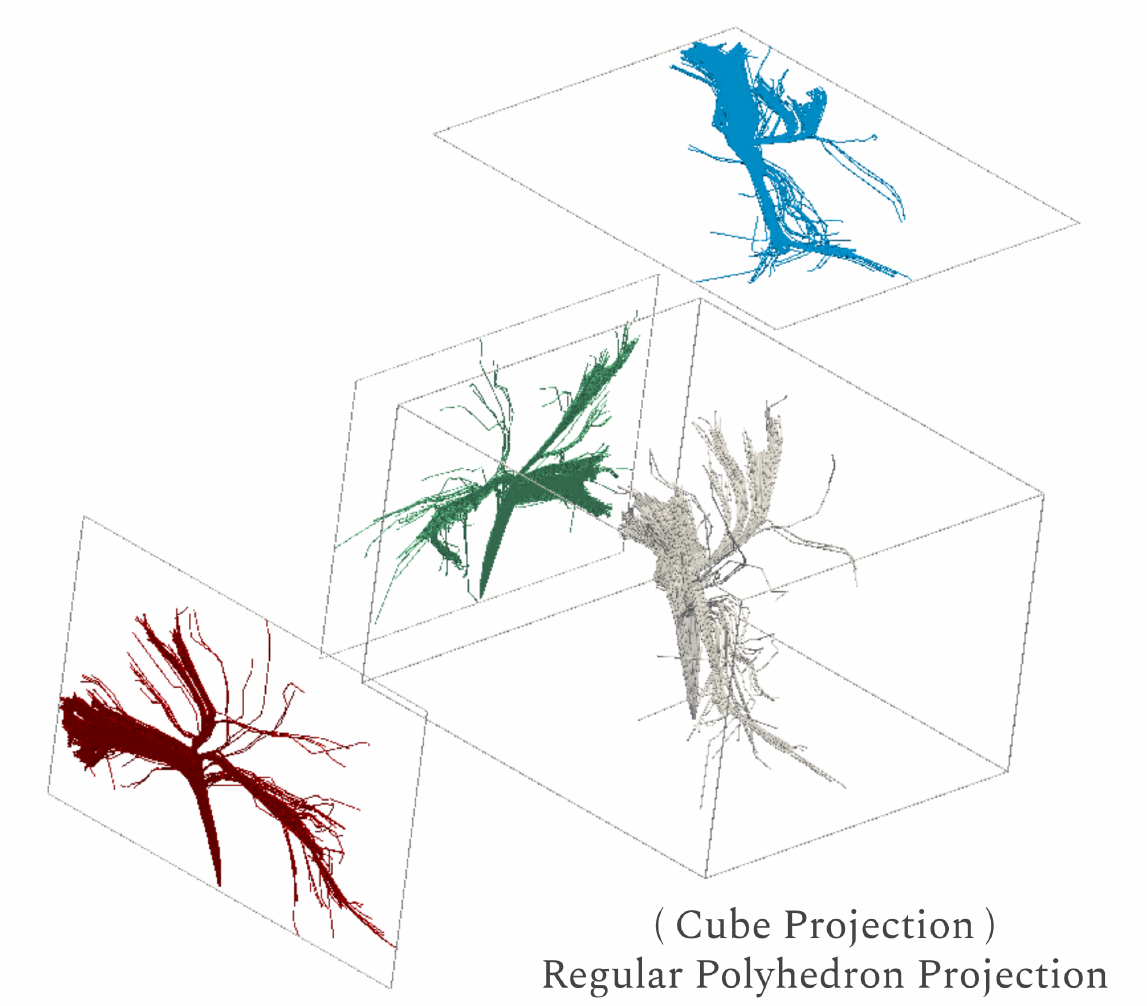}
	\caption{Regular polyhedron projection. Regular polyhedron is used to decompose the simplified streamlines. With cube as an example, the streamlines are projected onto three planes. The high order of polyhedron leads to in-depth decomposition. }
	\label{fig:RegularPolyhedronProjection}
\end{figure}

\begin{figure}[h]
	\centering
	\includegraphics[width=2.6in]{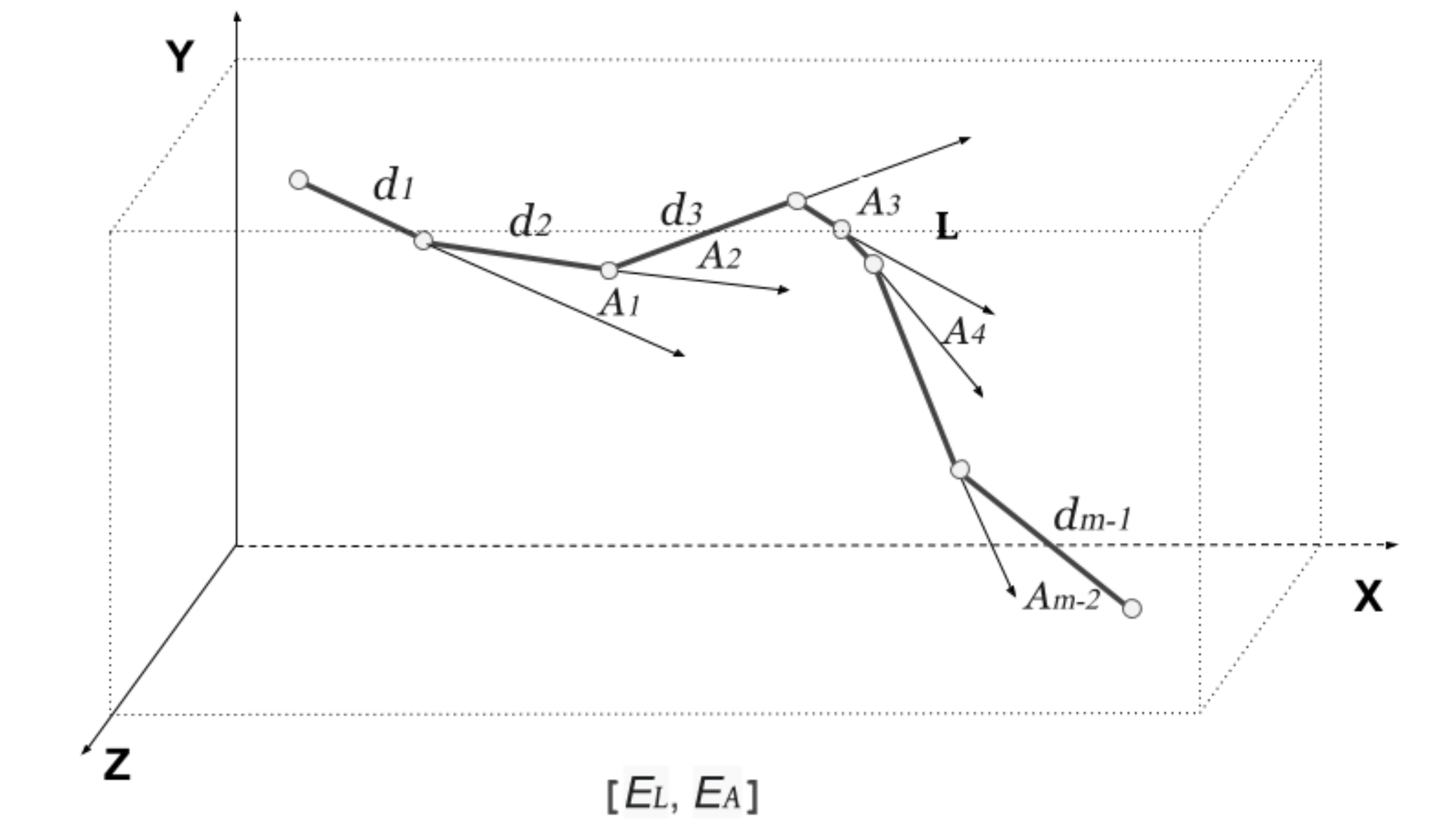}
	\caption{ Parameters used to compute entropy features. The streamline is depicted in 3D space, and adjacent distance ($d_i$) and joint angle ($A_i$) are shown. The number of distance is one more value than angles. }
	\label{fig:entropyFeature}
\end{figure}

\section{Feature Construction}

This section aims to construct streamline features that are inputs of the clustering algorithm. Prior to feature construction, we project the simplified streamlines to each plane of a regular polyhedron (Fig.\ref{fig:RegularPolyhedronProjection}). Thus, streamlines can be decomposed into more dimensions in physical space, which is conducive to the construction of more abundant features. The constructed features can reveal the spatial properties of a streamline adequately. The features are mainly classified into two categories: 3D space features and projection plane features.

\subsection{3D space features}
The 3D space features (entropy features) include \textit{linear entropy} ($E_L$) and \textit{angular entropy} ($E_A$). Proposed by Furuya and Itoh \cite{furuya2008streamline}, \textit{linear entropy} ($E_L$) is defined as follows:
\begin{equation}
	E_L = -\frac{ 1 }{\log_2 (m)} \sum_{i=1}^{m-1} \frac{d_i }{L_s}  {\log_2 \frac{d_i }{L_s}}
	\label{equation:EL}
\end{equation}
where $m$ is the number of the streamline nodes. As shown in Fig \ref{fig:entropyFeature}, $d_i$ is the length of $i-th$ segment and $L_s$ is the sum of $d_i$ values (the total length of the streamline). This feature can be used to discriminate streamline length segment properties but lacks the understanding of streamline angular properties. Thus, we introduce \textit{angular entropy} ($E_A$), which is proposed by Marchesin et al.\cite{marchesin2010view}, to this paper. This parameter quantifies the angular variation of a streamline and is computed by the following equation:
\begin{equation}
	E_A = -\frac{ 1 }{\log_2 (m-1)} \sum_{i=1}^{m-2} \frac{A_i }{L_A}  {\log_2 \frac{A_i }{L_A}}
    \label{equation:EA}
\end{equation}
where $m$ is the number of streamline nodes. $A_i$ is the absolute value of the angle between $i$-th vector and $i+1$-th vector. $L_A$ is the sum of the absolute value of $A_i$. Similar to \textit{linear entropy}, it quantifies the angular variation of a streamline rather than the curvature and is used for detecting flow phenomena.

 \begin{figure}[h]
 	\centering
 	\includegraphics[width=3.3in]{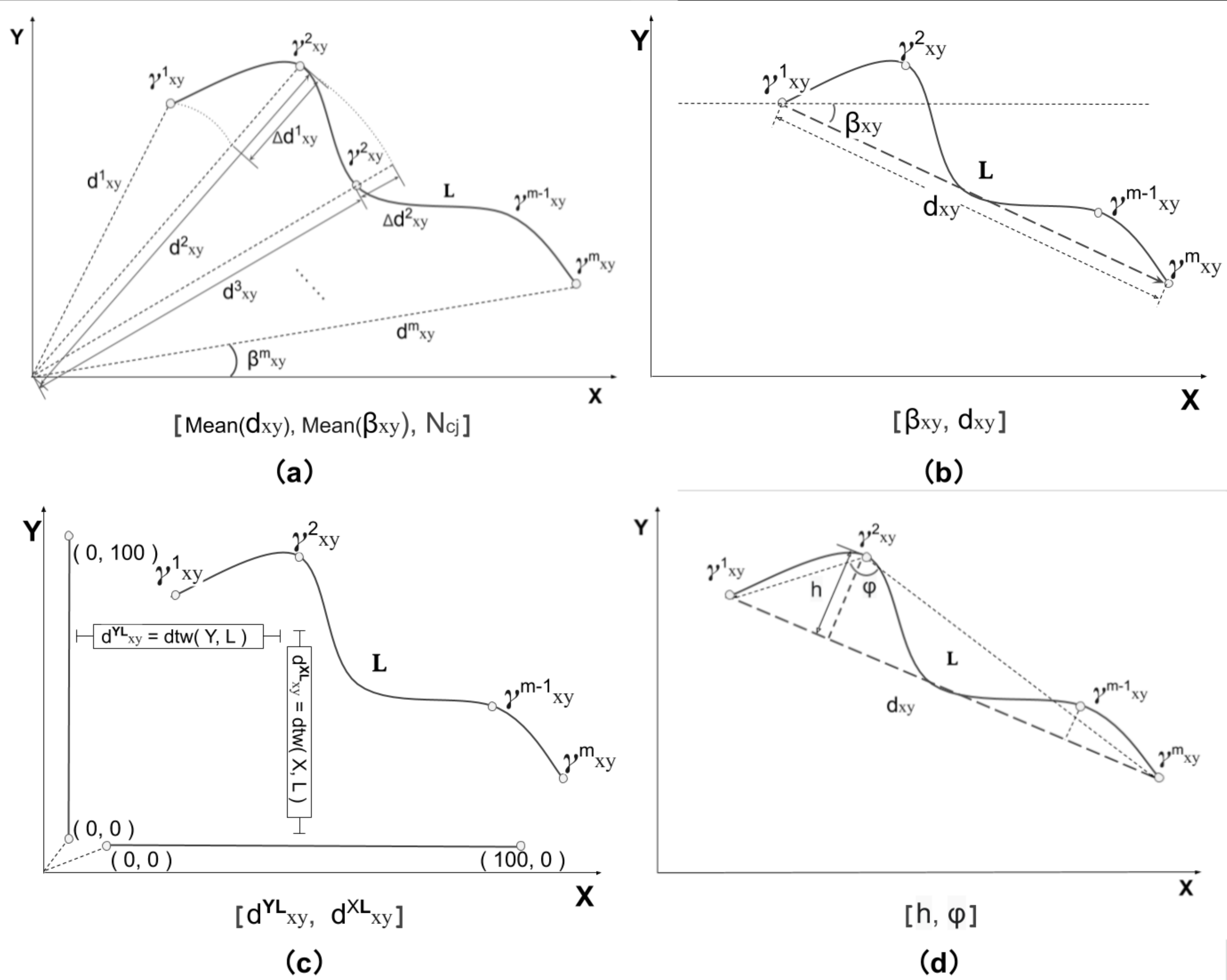}
 	\caption{Features in $XY$ plane. Projection plane features fall into four categories. (a) Drift features, which contain mean distance to the coordinate origin [$Mean(d_xy)$], mean angle to $x$ axis [$Mean(\beta_xy)$], and streamline drift rate ($N_{cj}$). (b) Orientation features shows the global streamline direction ($\beta_{xy}$) and absolute streamline length ($d_{xy}$). (c) Location features include the DTW distance of the streamline to each axis ($d^{YL}_{xy}$ and $d^{XL}_{xy}$).(d) Flexure features use the angle of the peak node to the first node and the last node ($\varphi$) and the distance of the peak node to the absolute streamline ($h$) to reflect the winding status of the streamline. }
 	\label{fig:features_v2}
 \end{figure}

\subsection{Projection plane features}
Projection plane features illustrate vector shape properties of a streamline on each plane. The projected curves of 3D streamlines that have different trends might be different. Using these projection curves, we can depict the shape properties of a streamline. With such a conjecture, we first project 3D streamlines to three planes($XY$, $YZ$, $ZX$). On each plane, we compute streamline features that fall into four categories: drift feature, orientation feature, location feature, and flexure feature. Fig ~\ref{fig:features_v2} shows how to construct these features in $XY$ plane. The projected streamline ($L$) on $XY$ plane consists of a list of nodes ($\gamma^1_{xy}, \gamma^2_{xy}, ..., \gamma^m_{xy}$). 

\textit{\textbf{Drift features}}. 
The streamline drift rate helps characterize the monotonicity of a streamline curve. As shown in Fig~\ref{fig:features_v2}(a), each node of the streamline is linked to the coordinate origin. The distance between a node and coordinate origin is described as $d^i_{xy}$ , and the corresponding angle of the node is $\beta^i_{xy}$. The mean value of the distances and angles are used to reduce dimensions of the streamline to a point in this plane. In addition, inspired by sequences and reversals test\cite{cowles1937some}, which is initially used to depict stock changes, we develop streamline drift rate ($N_{cj}$) to this paper, describing the monotonicity(how curve changes) of the streamline. Initially, we compute log gain rate ($I_i$) at each node:
\begin{equation}
I_i=\left\{
	\begin{array}{lcr}
	1, &  if & \ln(d^i_{xy}) - \ln(d^{i-1}_{xy}) \textgreater 0 \\ 
	0, &  if & \ln(d^i_{xy}) - \ln(d^{i-1}_{xy}) \leq 0  \\
	\end{array} 
\right.
\end{equation}
where, $i$ ranges from $0$ to $m$. $I_i > 0$ indicates that the streamline is monotonically increasing at $i$-th node, while $I_i < 0$ indicates that the streamline is monotonically decreasing.

With this transformation, a streamline can be transformed into a sequence of $0$ and $1$. In such a sequence, if any two adjacent numbers are 0 or 1, then they are called a sequence; on the contrary, if any two adjacent numbers are 0 and 1, or 1 and 0, then they are called a reverse. Streamline drift rate $N_{cj}$ is then computed based on streamline sequences and reversals:
\begin{equation}
\left\{
\begin{array}{l}
N_s = \sum_{i=1}^{m-1} Y_i \\ 
Y_i = I_iI_{i+1} + (1-I_i)(1 - I_{i+1} ) \\
N_r = m - N_s \\
N_{cj} = \frac{N_s}{N_r}
\end{array}
\right.
\end{equation} 
When the streamline is obviously monotonic, $N_{cj}$ should be significantly away from $1$. If the streamline is not monotonic, then $N_{cj}$ should gradually approach $1$.

\textit{\textbf{Orientation features}}.
Streamline orientation features help depict the overall trend of a streamline and contains two portions: the global streamline direction ($\beta_{xy}$) and absolute streamline length ($d_{xy}$), as seen in Fig~\ref{fig:features_v2}(b). The vector starts from the first node $\gamma^1_{xy}$ and ends with the last node $\gamma^m_{xy}$ is used to compute the orientation feature. $d_{xy}$ is the absolute value of this vector; and $\beta_{xy}$ is the angle of this vector and $X$ axis. 

\textit{\textbf{TLocation features}}.
Location features show the specific spatial position of the streamline at each plane. DTW\cite{giorgino2009computing} is a widely used distance calculation method that computes the optimal distance between points of two lists. By applying this method, we can obtain the distance between a streamline and each axis. ${d^{YL}}_{xy}$ and ${d^{XL}}_{xy}$ in Fig~\ref{fig:features_v2}(c) are the DTW distances of the streamline $L$ with $Y$ axis and $X$ axis, respectively. 

\textit{\textbf{Flexure features}}.
Flexure features show the winding status of a streamline rather than monotonicity. Fig.~\ref{fig:features_v2}(d) shows the schematic of these features. The dotted line that links the first node and the end node is used to compute the peak node, the one that has the maximum distance to the dotted line. The maximum distance ($h$) indicates the flexure variation range. $\varphi$ is the angle of the peak node vertex of a triangle that is formed with the first node, the peak node, and the last point.

\begin{equation}
L = \alpha L_r + (1 - \alpha) L_c
\label{eqn:L}
\end{equation}

\begin{equation}
\begin{cases}

\mathop{q_{ij}}=\frac{{{(1+{{\left\| \mathop{z}_{i}-\mathop{x}_{j} \right\|}^{2}})}^{-1}}}{\sum\nolimits_{j}{{{(1+{{\left\| \mathop{z}_{i}-\mathop{x}_{j} \right\|}^{2}})}^{-1}}}} \\

\mathop{p_{ij}}=\frac{{\mathop{q}_{ij}^{2}}/{\sum\nolimits_{i}{\mathop{q}_{ij}}}\;}{\sum\nolimits_{j}{{\mathop{q}_{ij}^{2}}/{\sum\nolimits_{i}{\mathop{q}_{ij}}}\;}} \\

\mathop{L_c}=KL(P||Q)=\sum\limits_{i}{\sum\limits_{j}{{{p}_{ij}}\log \frac{{{p}_{ij}}}{{{q}_{ij}}}}} \\

\mathop{L_r}=\sum\limits_{i=1}^{n}{\left\| \mathop{x_i}-\mathop{\tilde{x}_i} \right\|}_{2}^{2}
\end{cases}
\label{eqn:L1}
\end{equation}

\begin{figure}[h]
	\centering
	\includegraphics[width=2.6in]{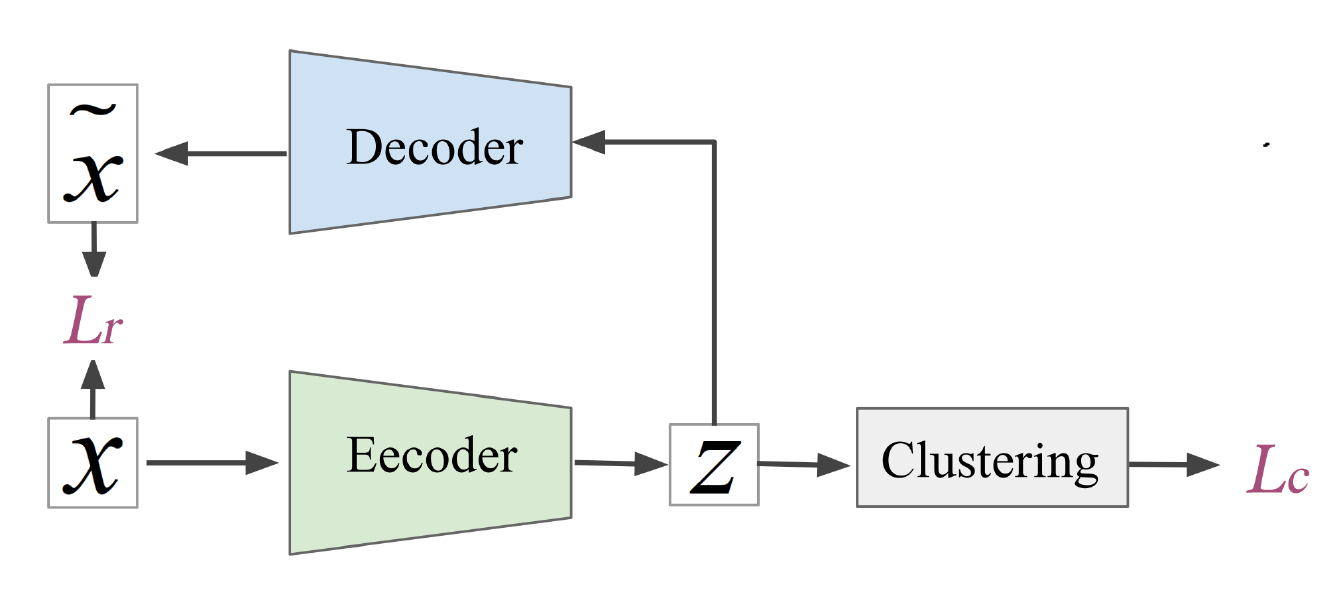}
	\caption{ Framework of deep embedded clustering. Multilayer encoder network initially maps the streamline features from feature space ($x$) to embedded space ($z$), while the decoder network produces reconstruction ($\tilde{x}$). Reconstruction loss ($L_r$) has been calculated by $x$ and $\tilde{x}$, while clustering loss ($L_c$) can be captured after performing clustering to embedded point $z$.}
	\label{fig:deepclustering}
\end{figure}

\section{Deep Clustering}
After feature construction, each streamline can be presented as a list that consists of thirty-one features. When the number of streamline increases, a deep neural network would be a good choice to handle the classification task, especially when the number of streamlines increases to thousands or millions. This deep neural network is also known as deep clustering.  

The IDEC algorithm has been proposed and applied to multiple datasets. Given that this clustering method can perserve the local structure of the data, we extend the scope of its application to streamline classification and empirically evaluate the proportion between reconstruction loss($L_r$) and clustering loss($L_c$). Intrinsically, our method is IDEC that considers of reconstruction loss effects and clustering loss effects on brain fiber data.

Fig.~\ref{fig:deepclustering} shows the framework of the clustering algorithm. A multilayer deep autoencoder has been used for classification, in which the encoder narrows down features from high-dimensional space (feature space) to low-dimensional space (embedded space) and the decoder reconstruct the embedded features to the original feature space. The reconstruction loss calculated between the original features and reconstructed features helps preserve the local structures of data gathering distribution, while clustering loss defined by DEC\cite{DEC} contributes to optimizing the deep neural network and updating cluster centers. By integrating the reconstruction loss and clustering loss, IDEC clustering can jointly optimize clusters while preserving the local data structure. However, the reconstruction and clustering loss are inherently contradictory, which is manifested in similarities and variances within and between clusters. The clustering loss targets at keeping within-cluster similarities and between-cluster variances while breaking the between-cluster similarities and within-cluster variances. By contrast, the reconstruction loss aims at protecting all the similarities and variances. However, the correlation between clustering loss and reconstruction loss is not purely negative. A scientific study should be conducted in such a case. In this paper, we only use linear negative correlation to estimate their correlations and evaluate the coefficients of losses in the current dataset.

Equation \ref{eqn:L} defined the objective, where $L_r$ and $L_c$ are the reconstruction loss and clustering loss, respectively. $\alpha$ is the coefficient used to balance the two losses. An empirical evaluation of $\alpha$ will be conducted in the next section. Equation \ref{eqn:L1} explains how to construct $L_r$ and $L_c$.  $q_{ij}$ is a t-distribution that maintains the data similarity and distance relations when reducing the dimensions from feature space($x_i$) to the embedded space($z_i$). $p_{ij}$ is then defined by $q_{ij}$ to form a self-training that is optimized by clustering loss($Lc$), also known as KL divergence loss, which is an asymmetric measurement to keep the value of $p$ and $q$ as close as possible. $L_r$ is the Mean Squared Error (MSE) that is computed by the input feature and the output of the autoencoder. The detailed description can be found in DEC\cite{DEC} and IDEC \cite{ijcai2017-243}. In the next section, we will introduce how we determine the parameters of deep clustering in the brain fiber clustering scenario.

\begin{figure*}[h]
	\centering
	\includegraphics[width=7in]{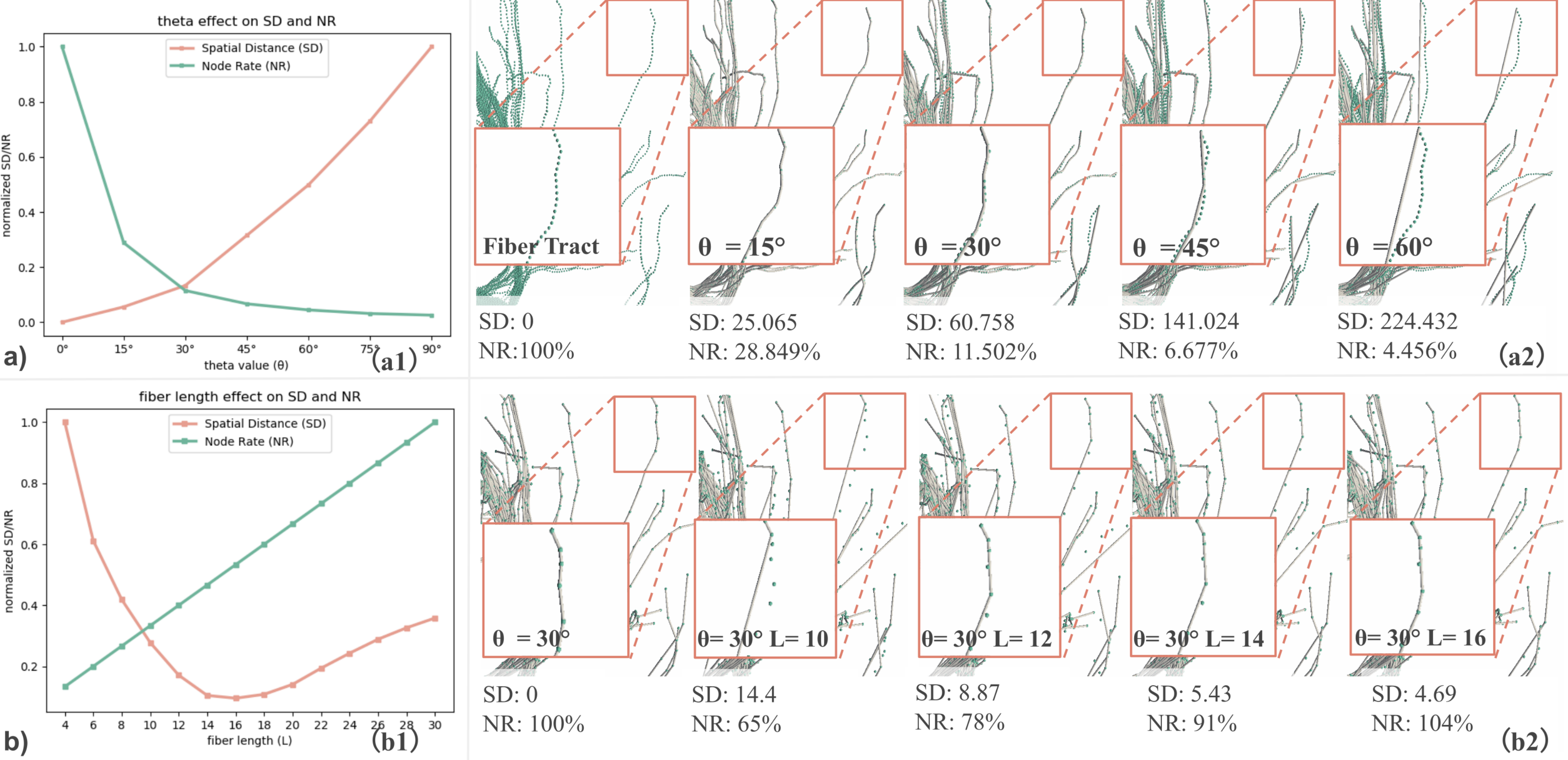}
	\caption{Effects of $\theta$ and $L$ on $SD$, $NR$, and brain fiber tracts. Row a) and row b) show the effects of $\theta$ and $L$, respectively. (a1) and (b1) illustrate how $SD$ and $NR$ changes with increasing values of $\theta$ and $L$, while (a2) and (b2) demonstrate the physical changes in the fiber tract with the two parameters. The green dotted lines illustrate the original brain fiber tracts, while the grey solid lines show the simplified brain fiber tracts. }
	\label{fig:Theta_FiberLength}
\end{figure*}

\section{Determination of Parameters}
\label{ParameterDetermination}
In this section, we employ an empirical approach used to determine the parameters of the pipeline for clustering brain fiber tracts. First, we define two indicators to determine the key parameters for brain fiber simplification. We then determine the two key parameters in view of the observations. Afterward, we determine the parameters in the deep clustering stage according to literature reviews and the results of experimental studies on brain fiber tracts. 

\subsection{Determination of Fiber Simplification Parameters}

The key parameters of fiber simplification are the theta value ($\theta$) and the simplified fiber length ($L$), which determine the curving similarity (How similar is the original fiber tracts to the simplified fiber tracts) and streamline compression degree (How many nodes of the original fiber tracts have been kept in the simplified fibers), respectively. In order to better understand the effect of $\theta$ and $L$ on fiber tracts and obtain suitable values for the two parameters, we defined two indicators ($SD$ and $NR$) and perform an empirical analysis on fiber tracts. 

The first indicator is the spatial distance ($SD$), which is computed by fastdtw \cite{salvador2007toward}, an effective distance calculation method for measuring the similarity between two sequences. It is used to compare the spatial similarity between original fiber tracts and simplified fiber tracts:
\begin{equation}
SD = \frac{ \sum_{i=1}^{M} fastdtw(L_i, S_i) }{M} 
\end{equation}
$M$ is the total number of the fiber tracts, $L_i$ and $S_i$ indicate the $i$-th fiber tract of the original fiber tracts and simplified fiber tracts, respectively. $fastdtw(L_i, S_i)$ calculate the distance between the two fiber tracts. A high $SD$ value demonstrates low similarity. 

The second indicator is the node rate ($NR$), which is used to calculate the compression degree of the simplified fiber tracts. $NR$ can be computed as follows: 
\begin{equation}
NR = \frac{ \sum_{i=1}^{M} C(S_i) }{ \sum_{i=1}^{M} C(L_i) } 
\end{equation}
$C(S_i)$ means node counts of $i$-th simplified fiber tract, while $C(L_i)$ shows node counts of $i$-th original fiber tract. A high $NR$ value illustrates a high fiber tract compression degree.

$SD$ is a distance measurement which generally ranges from zero to hundreds, while $NR$ is the coverage rate of the simplified fiber nodes from the original fiber nodes, which ranges from zero percent to one hundred percent if the original fiber is used as a benchmark. To better visualize the trends of $SD$ and $NR$ and better understand the $\theta$ and $L$ effects, we normalized the $SD$ and $NR$ and plot the curves for direct comparison.

We pursue the best quality of fiber simplification, which means high fiber tract curving similarity (low $SD$ value) and a high fiber tract compression degree (low $NR$ value). Fig.~\ref{fig:Theta_FiberLength} (a) helps find out the optimum $\theta$ value. \begin{math}\theta = 0\end{math} means the fiber tracts have not been simplified, while large $\theta$ value indicates a high level of fiber simplification. As shown in Fig.~\ref{fig:Theta_FiberLength} (a1), $SD$ and $\theta$ seem positively correlated, oppositely, $NR$ and $\theta$ show negative correlation. With $\theta$ value increasing from $0$ to $90$, the intersection of the two curves is close to the position, where $\theta = 30$. Basing on this observation, we take this value as the optimum empirical $\theta$ value. Fig.~\ref{fig:Theta_FiberLength} (a2) shows the rendering of the source fiber tracts and the simplified fiber tracts. From left to right, the figure shows how the simplified fiber tracts move away from the source fiber tracts with increasing $\theta$ value. We can clearly see that the simplified fiber tracts match the source fiber tracts well when $\theta$ reaches $30$. Meanwhile, the $NR$ value indicates that the compression rate of the simplified fiber tracts is $11.502\%$. In determining $L$ value, we conducted a similar experimental investigation. Fig.~\ref{fig:Theta_FiberLength} (b) shows the statistical chart and the rendering of fiber tracts. This analysis is based on $theta = 30$. Fig.~\ref{fig:Theta_FiberLength} (b1) shows the curves of $SD$ and $NR$ with increasing $L$. As shown in this plot, $NR$ shows a positive linear correlation with $L$, while $SD$ shows a non-linear correlation. The $SD$ reaches the minimum value when $L$ reaches $16$. The fiber tract rendering view (Fig.~\ref{fig:Theta_FiberLength} (b2) demonstrates that the simplified fiber tract changes with different $L$ values. The gray solid line and the green dotted line tend to coincide when $L$ increases. According to the above observation and analysis, we believe that the two values ($\theta = 30$ and $L = 16$) are the optimal values of the brain fiber data.

\begin{figure*}[h]
	\centering
	\includegraphics[width=7in]{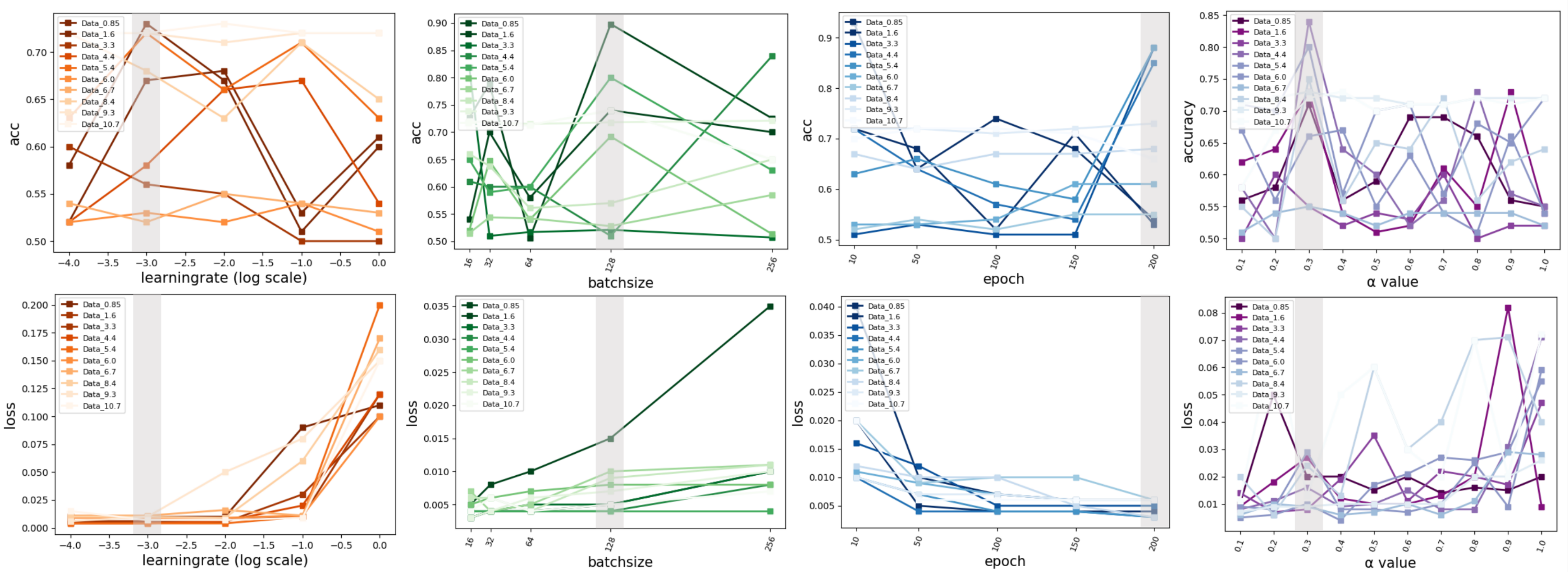}
	\caption{Performance of the optimization parameters on clustering accuracy/loss under $10$ brain fiber tract data. From left to right, the figure shows the effects of learning rate, batch size, epoch, and  $\alpha$ coefficient.}
	\label{fig:IDEC_Parameters}
\end{figure*}

\subsection{Determination of Deep Clustering Parameters}

The deep neural network is the representation learning component of deep clustering algorithms. The most widely used architectures are autoencoder based. In this paper, a multilayer deep autoencoder network is incorporated into the IDEC framework. Various of parameters need to be adjusted to train the deep learning model, including network parameters (number of layers, number of nodes, number of filters, and activation function,etc.), optimization parameters (batch size, number of the epoch, learning rate, etc.), and regularization parameters (weight attenuation coefficient, dropout ratio, etc.)
According to previous studies\cite{bottou2018optimization, goodfellow2016deep}, deep learning optimization is generally based on stochastic gradient descent (SGD). The updated rule of SGD highly relies on loss function, batch size, learning rate, and other suboptimal hyper-parameters\cite{smith2018disciplined, masters2018revisiting}. Classification accuracy performs. Thus, we mainly focus on tuning the optimization parameters to make it better-fit brain fiber tracts data, while other parameters are inherited from the study of IDEC\cite{ijcai2017-243}. Considering that unsupervised clustering accuracy and clustering loss are regularly used to evaluate the clustering performance\cite{yang2010image, min2018survey}, by evaluating optimization parameters on these two indicators, we expect to obtain optimum parameter values for the brain fiber tract data by evaluating the optimization parameters on the two indicators. 

A heuristic method is available for estimating a reasonable learning rate\cite{smith2017cyclical}. Researchers regard classification accuracy as a function of increasing learning rate. When plotting accuracy versus learning rate, one can easily choose an optimal learning rate for the test brain fiber dataset. Afterward, we estimated batch size and epoch for the brain fiber tracts. In general, grid search method would help obtain the optimal parameters within the specified parameter range. However, this method requires traversing the combination of all possible parameters, which is very time-consuming in the case of large data sets. The number of hyperparameters increases, and the computational complexity of the grid search will increase exponentially. Inspired by a previous study on tuning deep learning parameters \cite{masters2018revisiting}, the training of batch sizes and epochs have been performed over the same values of other (unadjusted) parameters, which corresponds to a constant computational complexity. We evaluated the performance of the parameters with $10$ brain fiber tract data, in which the fiber tract numbers range from $0.85K$ to $10.7K$.  

Fig.~\ref{fig:IDEC_Parameters} shows the performance of the optimization parameters on test accuracy/loss under the ten brain fiber tracts data. Curves with different transparency indicate brain fiber data that have different fiber numbers. The optimum parameters are shown in the gray zone of each subplot. The first column of Fig~\ref{fig:IDEC_Parameters} shows the effects of the learning rate. The learning rate for evaluation ranges from $0.0001$ to $1$, which is log scaled on the horizontal axis in this plot. The clustering loss consistently improves by increasing the learning rate, while the clustering accuracy varies a lot. However, we can see most of the brain fiber data to obtain peak clustering accuracy at the position of $log(learning rate) = -3.0$. The second column of Fig.~\ref{fig:IDEC_Parameters} shows the effects of batch size. As reported by \cite{kandel2020effect}, the typical batch size should be a power of $2$ that fits the memory requirements of the GPU or CPU. We start the batch size from $16$ until $256$. Some works in the optimization literature concluded that higher batch sizes usually do not achieve high accuracy and may lead to a significant degradation in the quality of the model\cite{radiuk2017impact, keskar2016large}. With this in mind, we ramp up the batch size for our model to see if we can obtain similar results. As shown in the second column of Fig.~\ref{fig:IDEC_Parameters}, the loss increases with batch size, and an optimal value exists for the batch size ($batch size = 128$), where the accuracies of most brain fiber data reach the maximum points. The third column of Fig.~\ref{fig:IDEC_Parameters} shows the effects of the epoch. We can see a pattern that the loss decreases with increasing the number of epoch from the overall view. A sharp decline appears when the number of epoch increase from $10$ to $50$, however, it drops slowly after that. Meanwhile, accuracy has no obvious trend in the plot view of epoch, and the change in accuracies differs from data. However, the accuracy of most data increases when the number of the epoch is over $150$. Thus, we regard $epoch = 200$ as the optimal value, which brings high accuracy and low loss for most of the brain fiber data. Finally, as shown in the last column of Fig.~\ref{fig:IDEC_Parameters}, we evaluate the performance of $\alpha$. Although both of the accuracy and loss differ greatly from the data, the figure also indicates the presence of an optimum value of $\alpha$. From an overall perspective, steep peaks appear at the position of $\alpha = 0.3$, which means that the accuracies of brain fiber data achieve the maximum values. The corresponding loss curves in the loss plot of $\alpha$ show relatively small values. Thus, we treat $\alpha = 0.3$ as the optimal ratio to balance $L_r$ and $L_c$. To sum up, our empirical study show that the optimization parameters ($learning rate = 0.001, batch size = 128, epoch = 200, and \alpha = 0.3$) would better fit the deep clustering of the brain fiber data. 

\begin{figure*}[h]
	\centering
	\includegraphics[width=7in]{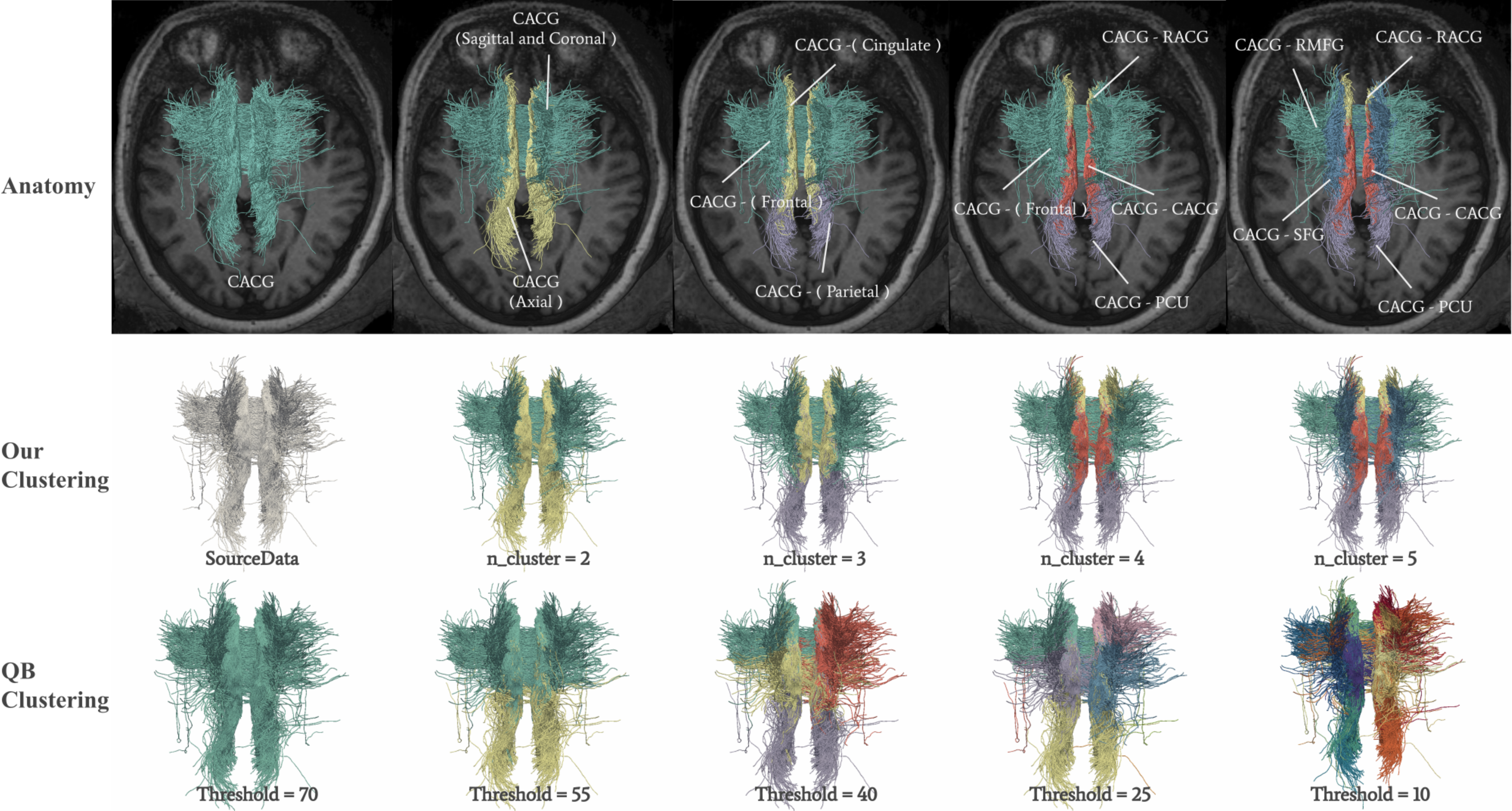}
	\caption{Qualitative comparison of the conventional rendering of CACG  brain fibers, our clustering results, and the QB clustering results.	}
	\label{fig:IDECvsQB_CACG}
\end{figure*}

\section{Clustering Results}
\label{clusteringresult}

In this section, we provide a visual comparison among the clustering results of our clustering method, QB fiber clustering method\cite{garyfallidis2012quickbundles}, and the brain anatomical structures. 
Before comparing the clustering results, we briefly introduce the difference between our clustering method and the QB clustering method in theory. QB clustering is an efficient spatial-distance based algorithm that can straightforward calculate the distances of brain fiber tracts. Given that the distance $threshold$ of the fiber tracts is the parameter of the QB clustering algorithm, the number of clusters in the clustering results is uncertain. This finding differs from data. The same $threshold$ may lead to a different number of clusters. Nevertheless, our clustering method highly relies on fiber curvature rather than pairwise spatial distance of the brain fiber tracts. We manually specify the number of clusters for each brain fiber data. This strategy will facilitate us to compare the clustering results directly. Brain anatomical structures are segmented based on one's cognitive function, which is of great value in clinical medicine research. 

Next, we briefly describe how to generate brain fiber tracts from the original MRI images and extract the brain fiber tracts of different functional regions. Meanwhile, we introduce the brain regions and brain lobes, as well as the abbreviations of brain regions that we used for comparison. Based on the generated brain fiber data, we provide an intuitive comparison among the three types of fiber tracts. We describe the comparison in detail at two levels (brain region level and whole-brain level).

\subsection{Brain Fiber Data}
Diffusion tensor imaging (DTI) and T1-weighted images were obtained to generate brain fiber tracts. After the image preprocessing, e.g., echo-planar Image/field map correction, eddy current correction, head motion correction, and B1 field inhomogeneity correction, we perform image registration, including intra-subject registration and inter-subject registration, using FLIRT \cite{greve2009accurate}, a fully automated robust and accurate tool for brain image registration. Intra-subject registration means the registration of different modality image data of an individual subject. Inter-subject registration maps the image data of different subjects in the same template, such as brain atlas. This would support brain functional region segmentation and direct comparison between different brains. The brain functional regions are segmented by using FreeSurfer\cite{FISCHL2012774}. After all these steps, we constructed brain fiber tracts by using MRtrix3 package \cite{TOURNIER2019116137}, a state of the art brain fiber tractography tool. The reconstruction steps are taking advantage of the Anatomically-Constrained Tractography (ACT) framework\cite{SMITH20121924}. With the default maximum angle between successive steps, it allows for neuronal axon reconstruction while retaining biological accuracy. While mapping brain atlas to whole-brain fibers, we can extract the brain fiber tracts of each brain functional region. The brain fiber tracts of a brain functional region usually connect to many other brain regions. Benefit from brain connectome components of MRtrix3 package, we could extract brain fiber tracts that connect among different brain functional regions. Brain regions in our paper are from the default FreeSurfer segmentation (desikan-killiany cortical atlas). The abbreviation of brain functional regions are as follows: Bankssts(BSTS), Caudal Anterior Cingulate(CACG), Lingual(LG), Rostral Anterior Cingulate(RACG), Precuneus(PCU), Rostral Middle Frontal(RMFG), Superior Frontal(SFG), Fusiform(FG), and Lateral Occipital(LOG). Meanwhile, the most commonly used brain lobes in medical research are the Frontal lobe, Parietal lobe, Occipital lobe, Temporal lobe, Cingulate lobe, Cerebellum lobe, and Subcortical lobe.

\subsection{Visual Comparison at Brain Region Level}
Two brain regions are used for evaluating our clustering results. One is the CACG brain region. We extract the region's brain fiber tracts from a subject and perform our clustering algorithm and QB clustering algorithms on it, as well as digging out the anatomy connections of the fiber tracts. The brain region's fiber tracts are used to compare and determine which clustering method would segment the brain fibers with more anatomical significance. The other is the BSTS brain region. We extract the regions' brain fiber tracts from multiple subjects and cluster the fiber tracts into several categories. Thus, we can see if the clustering method can produce results with consistency and robustness. 

Fig.~\ref{fig:IDECvsQB_CACG} illustrates the qualitative comparison of the conventional rendering of three types of CACG fiber tracts. The first row shows the anatomical structure of CACG brain fiber tracts. The CACG brain region is the brain region on the frontal part of the cingulate cortex, which not only links the left brain and the right brain but also many other connections among different functional regions. Thus, CACG brain fiber tracts can be divided into many subsets. Each of the subsets reveals the connection between different parts of the brain. Therefore, we labeled the fiber tracts with the abbreviations of brain regions at multiple levels. For instance, CACG(Sagittal and Coronal) indicates the transverse CACG brain fiber tracts that connect the left brain to the right brain; CACG-(Cingulate) means the fibers link the CACG region and Cingulate brain lobe; and CACG-PCU illustrates the fiber tracts that connect the CACG region and PCU region. The second row shows our clustering results of the CACG brain fibers that were classified into different categories, starting from the one category (source data) to five categories.  Our clustering result ($ncluster=2$) splits the CACG brain fibers into transverse fibers and longitudinal fibers. With the clustering number increases, it gradually segments CACG brain fibers into multiple subsets. As shown in Fig.~\ref{fig:IDECvsQB_CACG}, our clustering results and the fiber tract anatomy are visually of high similarity, although some small biases exist at the edge of each category. The third column shows the QB clustering results of the CACG brain fiber tracts. We test the clustering algorithm with many thresholds and finally select several threshold values (starting from $70$ and gradually decrease it in steps of $15$) for the fiber comparison. The fiber clustering results using the selected thresholds have a similar number of categories as our clustering results. Comparing the second plot and the third plot of the QB clustering results and the corresponding plots of our clustering results, we can clearly see that the QB clustering method clustered the fiber tracts spatially close into one category, while our clustering results maintain the fiber anatomical structures. In addition, with a decrease in $threshold$, the number of clusters of the QB clustering results increases rapidly. Users have difficulty in determining how many categories are present, which brings difficulties for further investigation of the brain fiber clustering results. To sum up, comparing with the QB clustering algorithm, our clustering method would better maintain the anatomical structures of brain fiber tracts at the brain region level.  

\begin{figure}[h]
	\centering
	\includegraphics[width=3.5in]{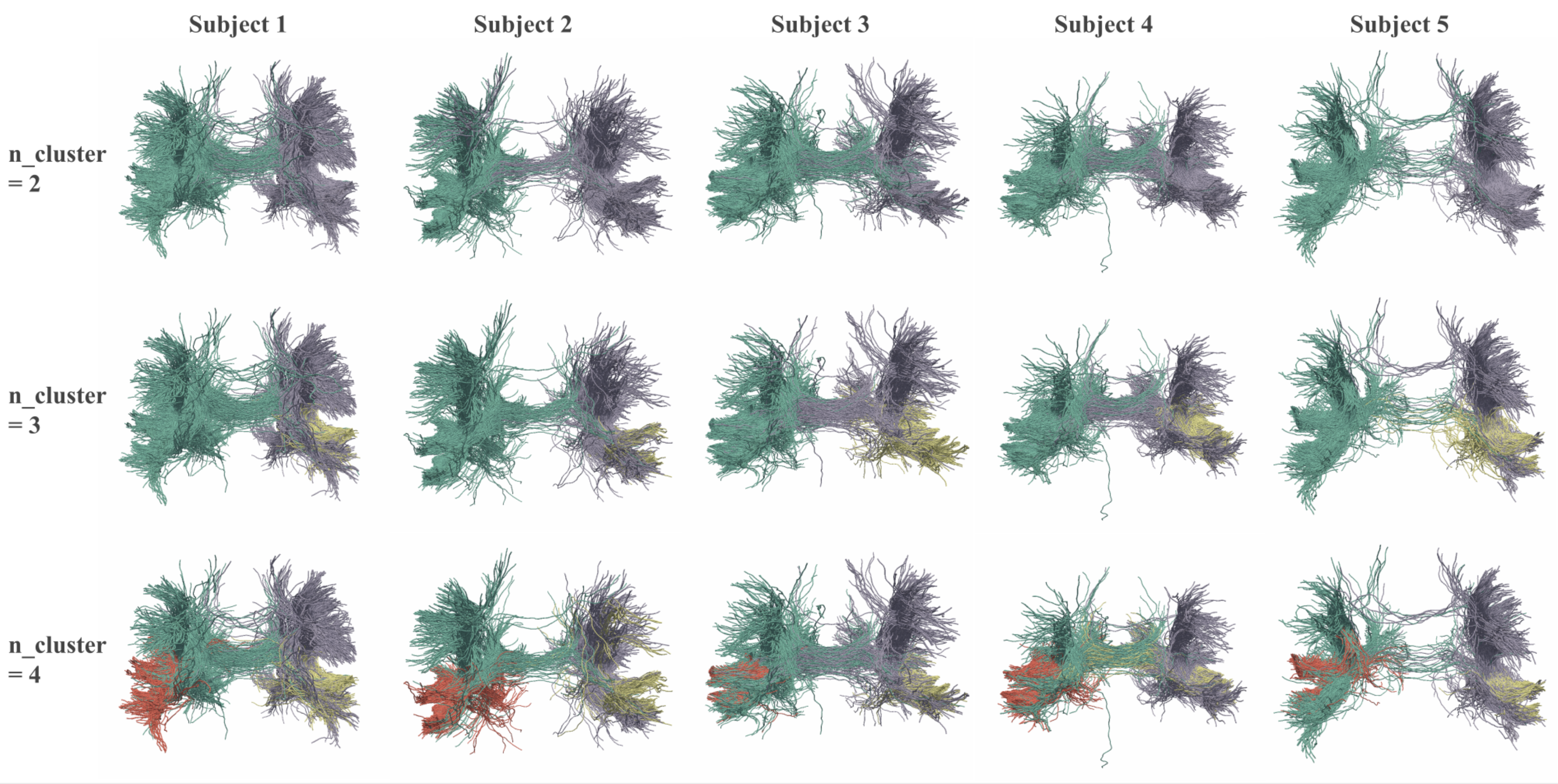}
	\caption{Clustering results of BSTS fibers in multiple subjects.}
	\label{fig:BSTS}
\end{figure}

In addition, we performed our clustering on multiple subjects. For each subject, we simply extract the fiber tracts of the BSTS brain region (the first brain region from the default FreeSurfer segmentation). We then cluster multiple times on each brain fiber tracts with successive cluster numbers. Fig.~\ref{fig:BSTS} shows our clustering results of BSTS fibers of multiple subjects. Given that each one's brain is unique, the extracted BSTS fiber tracts are not exactly the same. The fiber tracts have a similar shape on the whole but may differ in details, such as, fiber deformation and distortion. As shown in the first row, the BSTS fiber tracts were segmented into the left and right brain, when ($ncluster=2$). As the cluster number increases, the fiber tracts are divided into several groups, where the positioning of each group is relatively stable in view of the overall BSTS fibers. As shown in the third row of Fig~\ref{fig:BSTS}, the clustering results are symmetric on the whole and the fiber tracts with similar positions of each subject are segmented. Although we did not label the anatomical connections of each category, the clustering results of BSTS brain fiber tracts demonstrate the consistency of our clustering method.

\begin{figure*}[h]
	\centering
	\includegraphics[width=7in]{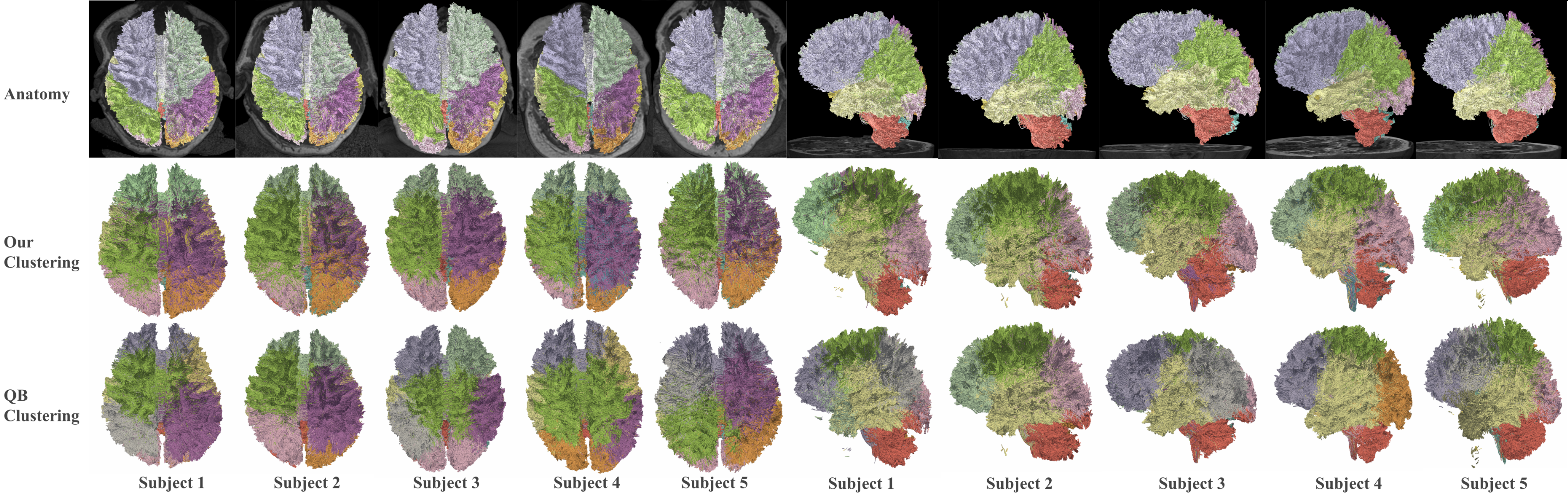}
	\caption{ The comparison of whole-brain fibers in axial and sagittal orientation among anatomical structures, our clustering method, and the QB clustering method.}
	\label{fig:IDECvsQB_WholeBrain}
\end{figure*}

\subsection{Visual Comparison at Whole-Brain Level}

We first derive whole-brain fiber tracts for multiple subjects. Each of the whole brains contains $100$ thousand fiber tracts. We then perform qualitative comparison among the QB clustering method, our clustering method, and brain functional mapping on the generated whole-brain fibers(Fig~\ref{fig:IDECvsQB_WholeBrain}). The left plots of Fig.~\ref{fig:IDECvsQB_WholeBrain} show the rendering of the three types of brain fiber tract segmentations at the axial direction, while the right plots of Fig.~\ref{fig:IDECvsQB_WholeBrain} show the brain fibers at the sagittal direction. In Fig.~\ref{fig:IDECvsQB_WholeBrain}, the first row shows the anatomy of brain fiber tracts, in which the brain fiber covers a slice of the MRI image and the fibers of each brain lobe has a unique color. The brain fiber tract segmentation using brain functional mapping has extremely high consistency. The second row and the third row of Fig.~\ref{fig:IDECvsQB_WholeBrain} show the rendering of our clustering results and QB clustering results, respectively. For better comparison, our clustering is with the parameter $n_cluster = 10$, while we set QB clustering parameter $threshold = 50$. This parameter value leads to the QB clustering results of whole-brain fibers at about $10$ categories, which sometimes may vary from $9$ to $11$. Fig.~\ref{fig:IDECvsQB_WholeBrain} shows an intuitive visualization of the clustering results of five subjects. Our clustering and QB clustering segment the whole brain fibers into multiple pieces; however, our clustering results seem more consistent on the brain fiber data. From the views of Fig.~\ref{fig:IDECvsQB_WholeBrain}, we can obviously see the fiber rendering of our clustering method and QB clustering method. Although some fibers are visually clustered incorrectly, a certain degree of similarity exists with the brain tracts that mapped with brain functional regions. On the one hand, it may reveal the internal correlation between brain fibers and brain functional areas, on the other hand, it gives a potential to create a robust brain fiber template that enables consistent anatomical segmentation. 

\begin{figure}[h]
	\centering
	\includegraphics[width=3.5in]{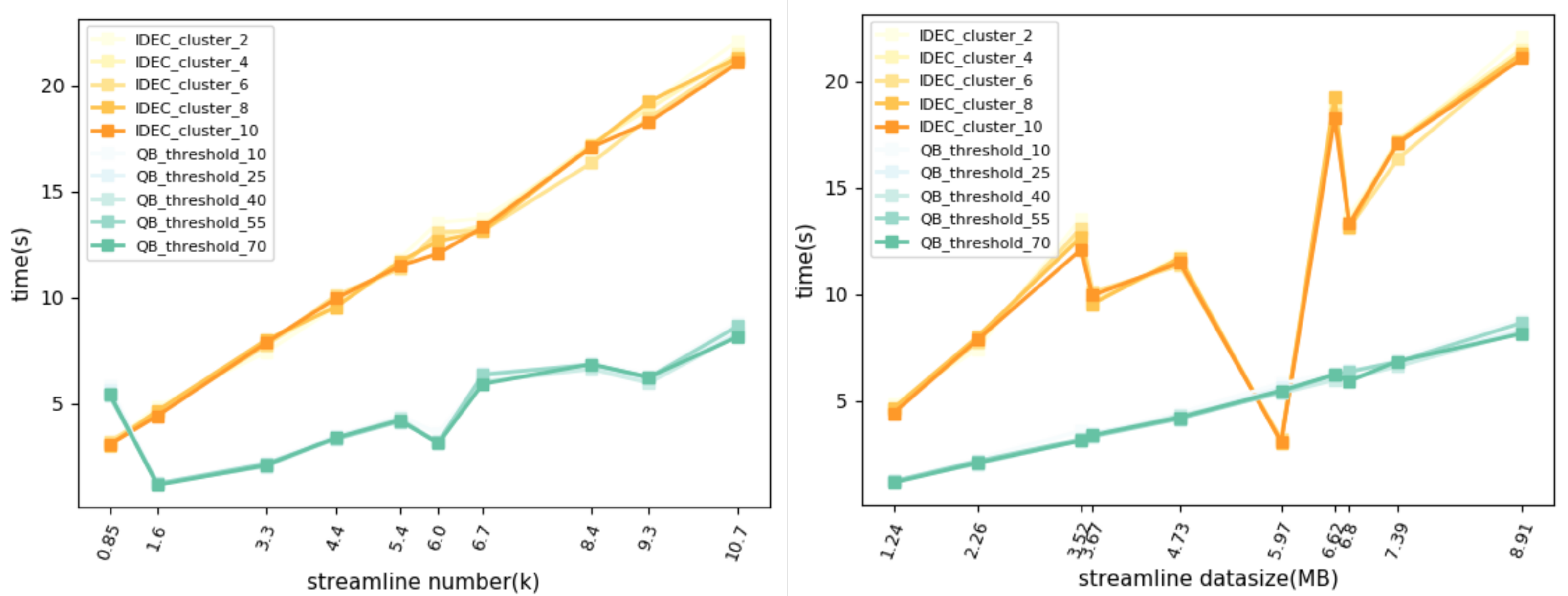}
	\caption{The performance of our clustering method and QB clustering method.}
	\label{fig:IDECvsQB_Times}
\end{figure}

\section{Performance and Evaluation}

Our clustering algorithm and QB clustering algorithm are implemented in the Windows laptop, which is configured as 4 Core Intel(R) Core(TM) i7-7700HQ CPU @ 2.80GHz, 16 GB memory, NVIDIA GeForce GTX 1070 graphics card, the development environment for Visual Studio Code. The clustering algorithms are programmed in Python. The deep clustering of our method is implemented with Keras TensorFlow 2.0\cite{bisong2019tensorflow}. Brain fiber rendering is performed using paraview\cite{ahrens2005paraview}, an open-source application for scientific visualization. 

Fig.~\ref{fig:IDECvsQB_Times} provides a quantitative comparison between our clustering method and QB clustering method. We evaluated the performance of the clustering algorithms with $10$ brain fiber tracts data. The fiber tract numbers range from $0.85K$ to $10.7K$; however, the data that have small fiber tract numbers may have large data size due to the different lengths of brain fibers. In Fig.~\ref{fig:IDECvsQB_Times}, the gradient orange lines indicate the time required of our clustering method under different number of clusters, which range from $2$ to $10$. The gradient blue lines indicate the time required of the QB clustering method underlying different $threads$ values, which range from $10$ to $70$. The left plot of Fig.~\ref{fig:IDECvsQB_Times} shows the performance of the two clustering methods on fiber tract numbers. From an overall perspective, the gradient orange lines are inclined upward when the fiber number gradually increases. Meanwhile, the gradient blue lines are presented as twisted lines. Hence, our clustering method is sensitive to fiber numbers in contrast to the QB clustering method not. The right plot of Fig.~\ref{fig:IDECvsQB_Times} shows the performance of the two clustering methods on brain fiber data size. In this plot, the gradient orange lines wrapped when the data size changes. By contrast, the gradient blue lines show a gradual upward trend when the data size increases. This finding indicates that QB clustering methods have a strong correlation with the data size.

The above observations demonstrate that the processing time of our method is positively related to the number of fibers, while the processing time of the QB clustering method has a positive correlation with data size. We should point out that our clustering method overall will be a bit more time-consuming than the QB clustering method because it takes about $20$ seconds to deal with brain fiber data that has $10K$ fiber tracts. Comparatively, it takes only about $8$ seconds using the QB clustering method. We also test the performance under the whole brain fiber data that have $100$ thousand fiber tracts. QB clustering method takes $1$ minute and our clustering method takes $3$ minutes. Considering our brain fiber clustering effects in section\ref{clusteringresult}, we regard that the efficiency of our clustering method would be acceptable. There's always a trade between efficiency and effects. 

\section{Discussion and Limitation}

We propose a fiber tract clustering framework for segmenting brain fiber tracts. These fiber tracts are generated from DTI images and presented as vector field streamlines. Each of the components of the framework (streamline simplification, feature construction, and deep clustering) is crucial to the clustering effects. The impact of streamline simplification is reflected in the accuracy of maintaining the original streamline shape. The influence of feature structure construction lies in the sufficiency of streamline property representation. The immediate impact of deep clustering is the accuracy of classification in feature space. Therefore, the parameters of the streamline simplification step and the deep clustering step as well as the order of regular polyhedron are vital to the effectiveness and efficiency of our clustering method. The parameter values are determined by qualitative observation and quantitative analysis of multiple brain fiber data. Such an empirical approach has a strong dependence on data, which leads to the lack of generalization ability of the parameters. However, the rendering of our clustering results on brain fiber data reveals that our method can better segment brain fiber tracts with anatomic significance than the QB clustering method. It takes seconds to run our clustering method on fiber tracts at the brain region level and minutes on fiber tracts at the whole-brain level. Overall, our clustering method can effectively and efficiently segment brain fiber tracts into meaningful substructures. One direction for further research is toward spatially invariant feature localization and extraction and extending our clustering framework to other research datasets.

Despite the successes of our work on brain fiber data, our framework and our study are not without limitations. 1) The efficiency of our clustering method is strongly affected by the streamline simplification algorithm, which is actually a greedy algorithm. An optimized and faster simplification method would improve the clustering efficiency. 2) Currently, we use a cube to decompose fiber tracts and construct brain fiber features, which limit the accuracy of features to the original fibers. The high order of polyhedron
would lead to more precise decomposition. 3) The IDEC clustering method inherently relies on 20-round kmeans for determining seed points, which brings some uncertainties and may yield slightly different clustering results. 

\section{Conclusion}
We provide a vector field streamline clustering framework that can effectively segment brain fiber tracts into pieces of substructures, which can help researchers maintain a sense of awareness about brain anatomical structures. The framework includes a vector field streamline simplification algorithm, a group of feature construction methods based on regular polyhedron projection, and an adaptive improvement of the IDEC clustering method. Qualitative and quantitative comparisons have been performed between our clustering method and QB clustering method using real brain fiber data. The results indicate that our approach has the potential to create a robust fiber bundle template and would benefit researchers by helping them effectively facilitate deeper investigation into brain fiber tracts. 

\section*{Acknowledgment}
This research is sponsored in part by the Nature Science Foundation of China through grant 61976075.


\ifCLASSOPTIONcaptionsoff
  \newpage
\fi



\bibliographystyle{abbrv-doi}
\bibliography{references}

\end{document}